\documentclass[3p,times,procedia]{elsarticle}
\usepackage{nupha_ecrc}

\volume{00}

\firstpage{1}

\journalname{Nuclear Physics A}

\runauth{I. Vitev}

\jid{nupha}

\jnltitlelogo{Nuclear Physics A}

\usepackage{amssymb}

\usepackage[figuresright]{rotating}

\begin{document}

\begin{frontmatter}



\%

\title{Soft-collinear effective theory for hadronic and nuclear collisions: The evolution of jet quenching from RHIC to the highest LHC energies}


\author{Ivan Vitev}

\address{Los Alamos National Laboratory, Theoretical Division, Mail Stop B283, Los Alamos, NM 87545, USA}

\begin{abstract}

In the framework of soft-collinear effective theory with Glauber gluons, results and predictions for inclusive  hadron suppression, based upon in-medium parton shower evolution, are presented for  Au+Au and  Pb+Pb collisions at RHIC and LHC energies $\sqrt{s}=200$~AGeV and   $\sqrt{s}=2.76, \,  5.1$~ATeV, respectively. The  $\rm SCET_G$ medium-induced splitting kernels are further implemented to evaluate the attenuation of reconstructed jet cross in such reactions  and to  examine their centrality and radius $R$  dependence. Building upon a previously developed method to systematically resum the jet shape at next-to-leading logarithmic accuracy,  a quantitative understanding of the jet shape modification measurement in Pb+Pb collisions at  $\sqrt{s}=2.76$~ATeV at the LHC can be achieved. Predictions for photon-tagged jet cross sections and shapes, that can shed light on the parton flavor dependence of in-medium parton shower modification, are also given.

\end{abstract}


\end{frontmatter}



\section{Introduction}
\label{intro}


Effective field theory (EFT) is a powerful framework based on exploiting symmetries and controlled expansions for problems with a natural separation of energy or distance scales. EFTs are particularly important in QCD and nuclear physics. An effective theory of QCD, ideally suited to jet applications, is soft-collinear effective theory (SCET)~~\cite{Bauer:2000ew}. Recently, first steps were taken to extend SCET and describe jet evolution in strongly-interacting matter~\cite{Idilbi:2008vm} 
by introducing a transverse momentum exchange Glauber gluon mode between the collinear partons and the medium quasi-particles.  The newly constructed theory, called soft-collinear effective theory with Glauber gluons ($\rm SCET_G$),
was used to derive all  $O(\alpha_s)$ $1\to 2$ medium-induced splitting kernels and discuss higher order $O(\alpha_s^2 )$ corrections to the  medium-modified jet substructure~\cite{Ovanesyan:2011kn}. These medium-induced splitting kernel are the universal key ingredients in the evaluation of a wide variety of hadronic and jet observables in nucleus-nucleus collisions. The theoretical advances reported here allowed us for the first time to go beyond the traditional energy loss approximation to parton propagation in matter and to unify the treatment of vacuum and medium-induced parton showers~\cite{Kang:2014xsa}. They provide quantitative control over the uncertainties associated with the implementation of the in-medium modification of hadron production cross sections and help accurately constrain the coupling between the jet and the medium.

 \section{Jet quenching from QCD evolution}
 \label{inclhadrons}

Elementary parton branching is the essential steps in the formation of a parton shower. It was demonstrated that in the ambiance of dense QCD matter
full splitting functions are equal to the sum of the vacuum  ones and the corresponding in-medium contribution:
\begin{eqnarray}
P_{i\rightarrow jk}^{\rm med}(x,{\bf k}_{\perp};\beta)=P_{i\rightarrow jk}^{\rm vac}(x)+P_{i\rightarrow jk}^{(1)}(x, {\bf k}_{\perp};\beta)\,,
\label{gensum}
\end{eqnarray}
where $x$ is the momentum fraction carried by the daughter parton, ${\bf k}_{\perp}$ is the transverse momentum relative
to the parent parton, and $\beta$ collectively describes the medium properties.
The splitting functions factorize from the hard scattering process and are gauge invariant. 
They govern the evolution of parton distribution functions and fragmentation functions (FFs) in the medium. Here, we are interested in final-state
interactions and the generalized DGLAP evolution equations for FFs in the medium are written down as follows~\cite{Kang:2014xsa}
\begin{eqnarray}
\frac{d D_{h/q}(z,Q)}{d \ln Q}&=&\frac{\alpha_s(Q)}{\pi}\int_{z}^1 \frac{d z'}{z'}\left[P^{\rm med}_{q\rightarrow qg}(z', Q; \beta)D_{h/q}\left(\frac{z}{z'},Q\right)
+P^{\rm med}_{q\rightarrow gq}(z', Q; \beta) D_{h/g}\left(\frac{z}{z'},Q\right)\right]\, ,
\label{eq:mAP10}
\\
\frac{d D_{h/g}(z,Q)}{d \ln Q}&=&\frac{\alpha_s(Q)}{\pi}\int_{z}^1 \frac{d z'}{z'}\Big[P^{\rm med}_{g\rightarrow gg}(z', Q; \beta)D_{h/g}\left(\frac{z}{z'},Q\right)
+P^{\rm med}_{g\rightarrow q\bar q}(z', Q; \beta)\sum_q D_{h/q}\left(\frac{z}{z'},Q\right)\Big]\, .
\label{eq:mAP30}
\end{eqnarray}
These evolution equations also encode the effect of parton energy loss from multiple soft gluon emissions.
In the $x \ll 1$ limit, the solution of Eqs.~(\ref{eq:mAP10}), (\ref{eq:mAP30}) is related to the traditional parton energy loss
picture~\cite{Kang:2014xsa}.

\begin{figure}[!t]
\begin{center}
\includegraphics[width=7.5cm]{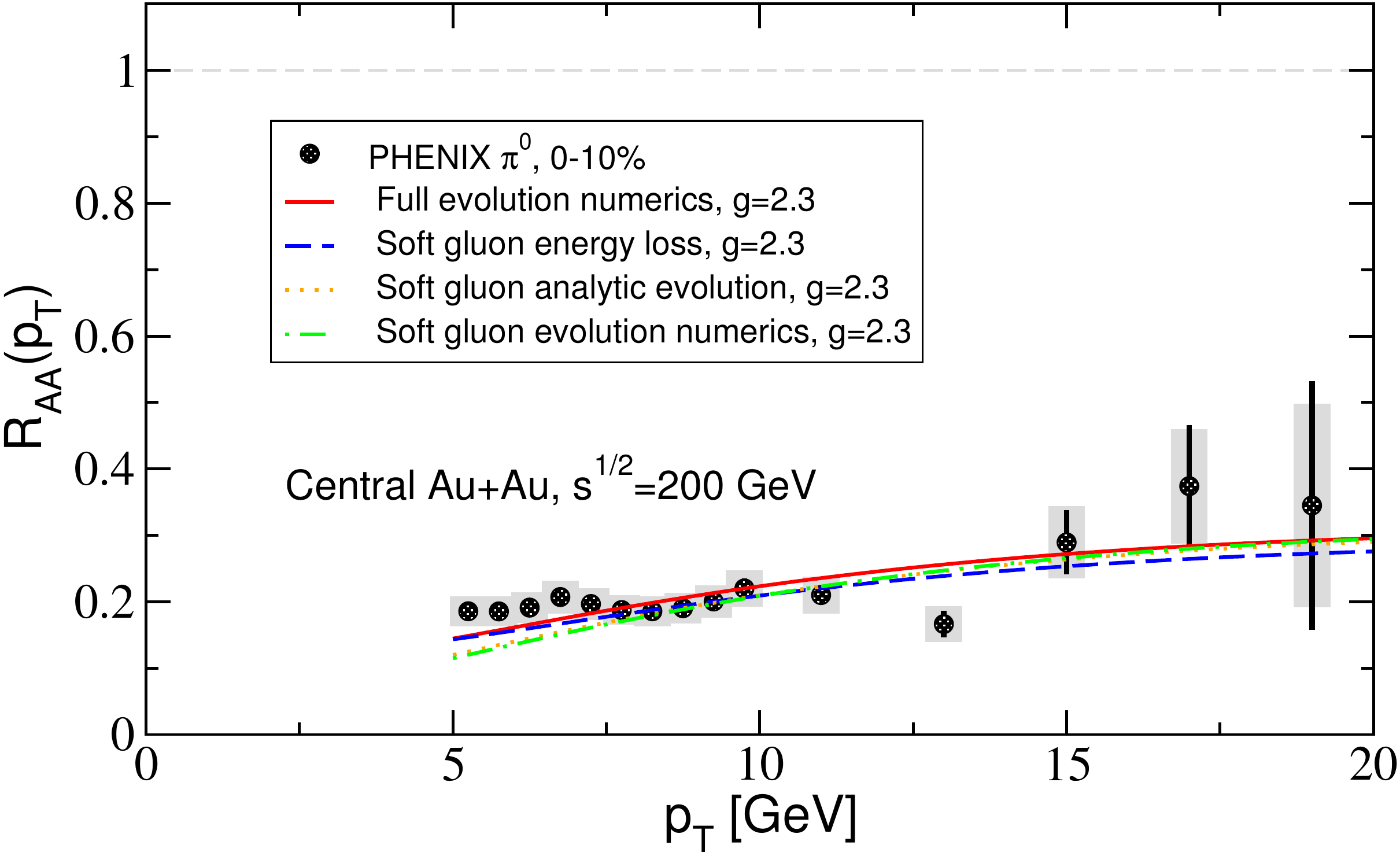} \ \ 
\includegraphics[width=7cm]{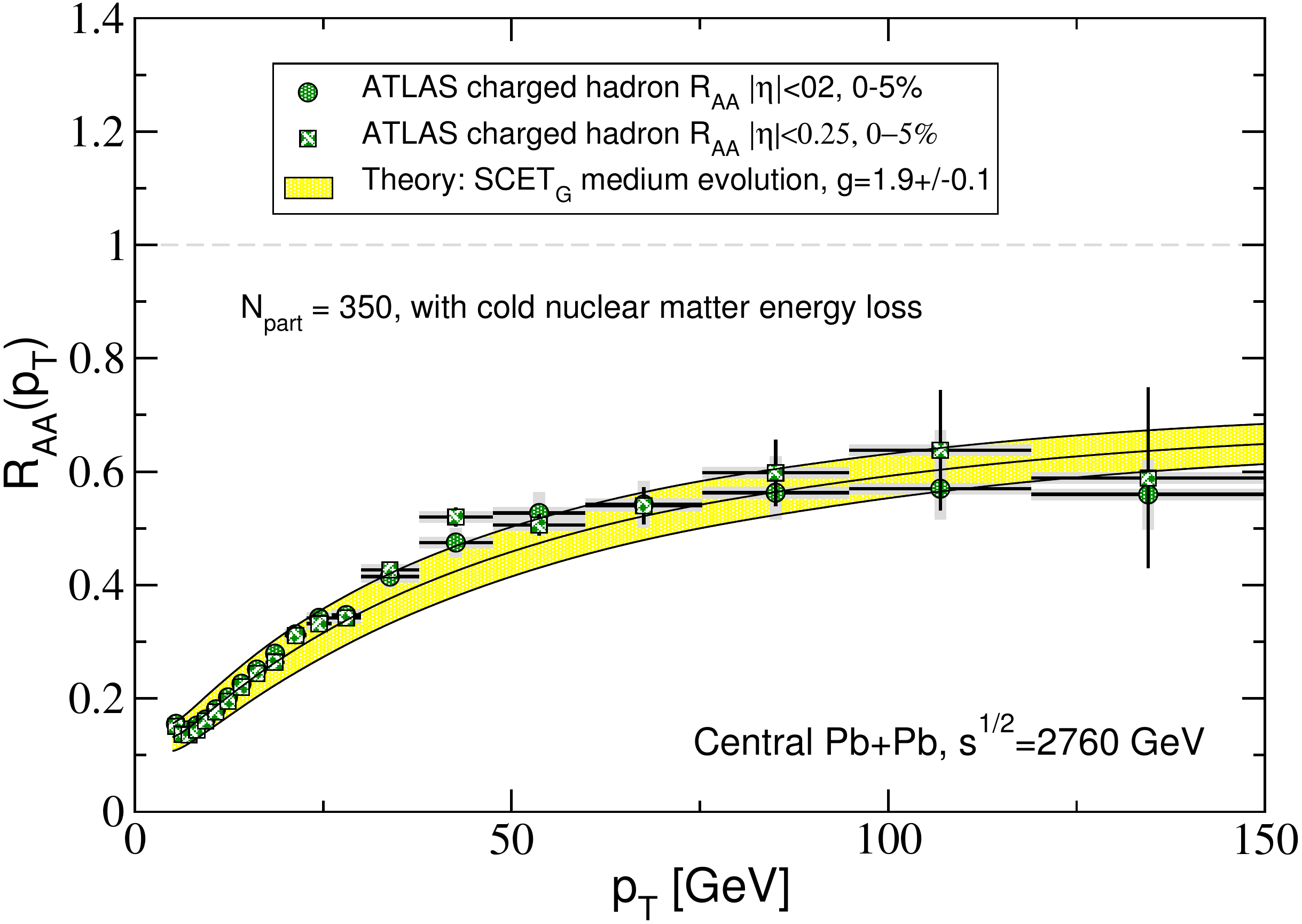}
\vspace*{-0mm}
\caption{ Left panel: four different calculations of $\pi^0$  $R_{AA}$ at RHIC with $g=2.3$ are compared to 
the PHENIX suppression measurements in central Au+Au collisions. 
Right panel:  nuclear modification factor $R_{AA}$ for charged hadrons is calculated in central Pb+Pb collisions at the LHC $\sqrt{s}=2.76$~ATeV, and compared with the ATLAS experimental data.
}\label{fig:rhicRAA}
\vspace*{-3mm}
\end{center}
\end{figure}

The medium-evolved FFs can be implemented in phenomenological calculations of hadron suppression in heavy ion collisions.
Comparison of the four different evaluations of the nuclear modification $R_{AA}$ for RHIC $\sqrt{s}=200$~AGeV in central Au+Au reactions is shown in the left panel of Fig.~\ref{fig:rhicRAA}. The calculations provide adequate description of the attenuation of the inclusive $\pi^0$ cross
section measured by the PHENIX experiment~\cite{Adare:2008qa}. The difference between the full solution
to the DGLAP evolution equations, semi-analytic solutions in the soft gluon limit, and the traditional energy loss approach are very small and corresponds to $\sim 5\%$ uncertainty in the determination of the coupling between the jet and the medium.  
Results for the suppression of inclusive charged hadron production at  $\sqrt{s}=2.76$~ATeV in central Pb+Pb collisions at the LHC are shown in the right panel of Fig.~\ref{fig:rhicRAA} and ATLAS data is included for comparison. Calculations based on solutions of the evolution equations with in-medium splitting kernels give good description of the centrality and transverse momentum $p_T$ dependence of charged hadron production at the LHC, while for neutral pions measurements to higher $p_T$ will help better assess if any discrepancy between theory and measurements exists~\cite{Abelev:2012hxa}.

\section{Jet cross sections and jet shapes from $\rm SCET_G$ }
 \label{inclhadrons}

At leading order, the jet energy function associated with parton $i$ with the collinear momentum $p$ splitting into $k=(x\omega,k_\perp^2/x\omega,k_\perp)$ and $p-k$ can then be written as follows,
\begin{equation}
    J_{\omega,E_r}^{i}(\mu)=\sum_{j,k}\int_{PS} dxd k_\perp {P}_{i\rightarrow jk}(x, k_\perp) E_r(x, k_\perp)\;,
\end{equation}
where ${P}_{i\rightarrow jk}(x, k_\perp)$ are the collinear parton splitting functions. Here,  $E_r(x, k_\perp)$ is the measurement function associated with the jet energy function. Since the splitting functions are directs sums of their vacuum 
and medium-induced components Eq.~(\ref{gensum}), we find
\begin{equation}
    J_{\omega,E_r}(\mu) = J^{vac}_{\omega, E_r}(\mu)+J^{med}_{\omega, E_r}(\mu)\, ,  \quad {\rm where} \quad
\frac{d J^{i}_{\omega,E_r}(\mu)}{d\ln\mu}
    =\left[-C_i\Gamma_{\rm cusp}(\alpha_s)\ln\frac{\omega^2\tan^2\frac{R}{2}}{\mu^2}-2\gamma^{i}(\alpha_s)\right]J^{i}_{\omega,E_r}(\mu) \, , 
\end{equation}
where $i=q, g$ with $C_q=C_F$ and $C_g=C_A$ the Casimir operators of the fundamental and adjoint representations in QCD, and $\Gamma_{\rm cusp}$ and $\gamma^i$ are the cusp and non-cusp anomalous dimensions~\cite{Chien:2014nsa}.
Furthermore, the in-medium splitting functions can be used to generalize the concept of energy loss for jets with 
reconstruction parameter $R$ beyond the soft gluon approximation. More specifically, the fractional medium-induced energy loss of quark and gluon jets is
\begin{eqnarray}
    \epsilon_q
    &=& \frac{2}{\omega}\Big[\int_0^{\frac{1}{2}} dx k^0
    +\int_{\frac{1}{2}}^1 dx (p^0-k^0)
    \Big]\int_{\omega x(1-x)\tan\frac{R}{2}}^{\omega x(1-x)\tan\frac{R_0}{2}}dk_\perp   \frac{1}{2} \Big[  {\cal P}^{med}_{q\rightarrow qg}(x,k_\perp) + {\cal P}^{med}_{q\rightarrow gq}(x,k_\perp)     \Big]    \;, \\
    \epsilon_g
    & = & \frac{2}{\omega}\Big[\int_0^{\frac{1}{2}} dx k^0
    + \int_{\frac{1}{2}}^1 dx (p^0-k^0)
    \Big]\int_{\omega x(1-x)\tan\frac{R}{2}}^{\omega x(1-x)\tan\frac{R_0}{2}}dk_\perp \frac{1}{2} \Big[
    {\cal P}^{med}_{g\rightarrow gg}(x,k_\perp)+   \sum_{q,\bar{q}}{\cal P}^{med}_{g\rightarrow q\bar q}(x,k_\perp)\Big].
\end{eqnarray}
Note that in the splitting $g\rightarrow gg$ the final-state partons are identical particles. Here $R$ is the angular parameter used in the jet reconstruction, and $R_0$ is of ${\cal O}(1)$ in QCD which sets the region of the use of collinear parton splitting functions.

\begin{figure}[!t]
\begin{center}
\includegraphics[width=7.4cm]{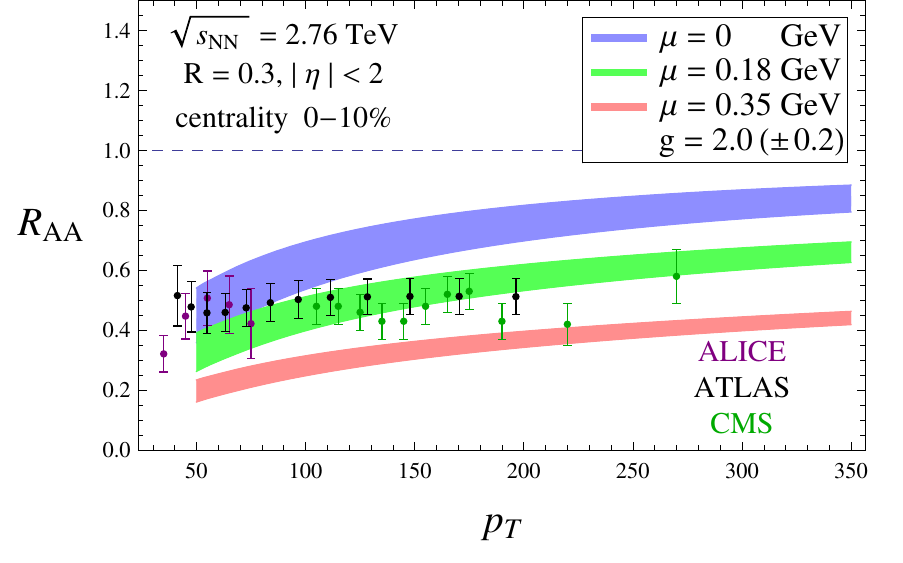} \ \ 
\includegraphics[width=7.4cm]{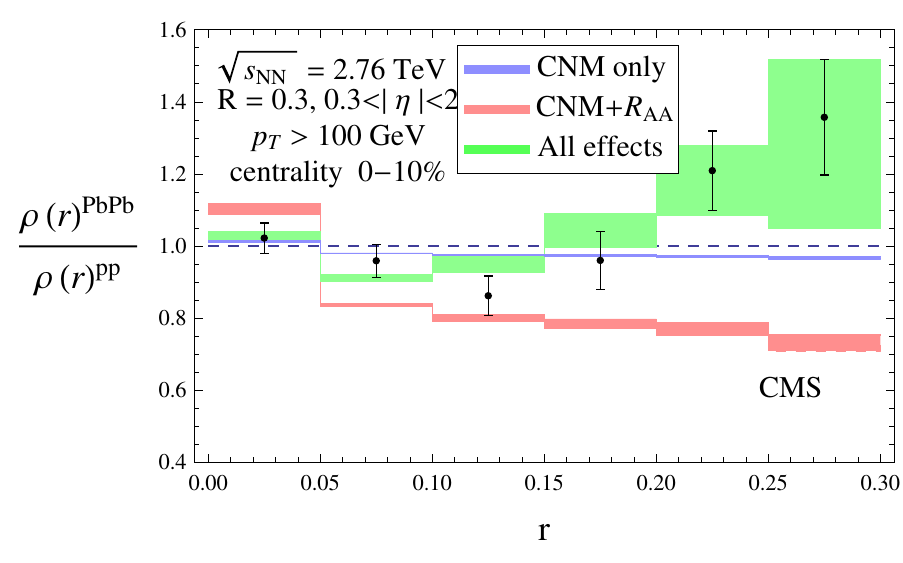}
\vspace*{-6mm}
\caption{ Left panel: theoretical calculations for the suppression of inclusive jets with $R=0.3$ as a function of the jet transverse momentum to experimental data in Pb+Pb collisions at $\sqrt{s}=2.76$~ATeV are compared to ALICE, ATLAS and CMS data.
Right panel:  modification of differential jet shapes of inclusive jets at the LHC. The modification is presented as the ratio of the jet shapes $\rho^{\rm PbPb}(r)/\rho^{\rm pp}(r)$ in Pb+Pb and p+p collisions.  The blue band corresponds to the calculations including only the CNM effects, the red band adds the jet-medium interaction but with the jet-by-jet shape modification turned off, and the green band correspond to the full calculation. Data is from CMS with $R=0.3$.
}\label{fig:jetMod}
\vspace*{-3mm}
\end{center}
\end{figure}

The results for the medium-modified jet energy functions and reconstructed jet energy loss can be implemented in phenomenology. The left panel of Fig.~\ref{fig:jetMod} shows the suppression of inclusive jet production in central Pb+Pb collisions at the LHC.
The bands correspond to couplings between the jet and the medium $g=2\pm0.2$ and no cold nuclear matter (CNM) energy loss, small CNM e-loss, and moderate CNM e-loss~\cite{Vitev:2007ve}. Experimental data is from the ALICE, ATLAS and CMS 
collaborations~\cite{Abelev:2013kqa}. In the right panel of Fig.~\ref{fig:jetMod} comparison of the calculated jet shape 
modification (green band) to CMS results at $\sqrt{s}=2.76$~ATeV is shown.  The contribution of various initial-state and final-state effects is also presented.

\section{Conclusions}
\label{Conclude}

\begin{figure}[!t]
\begin{center}
\includegraphics[width=6.2cm]{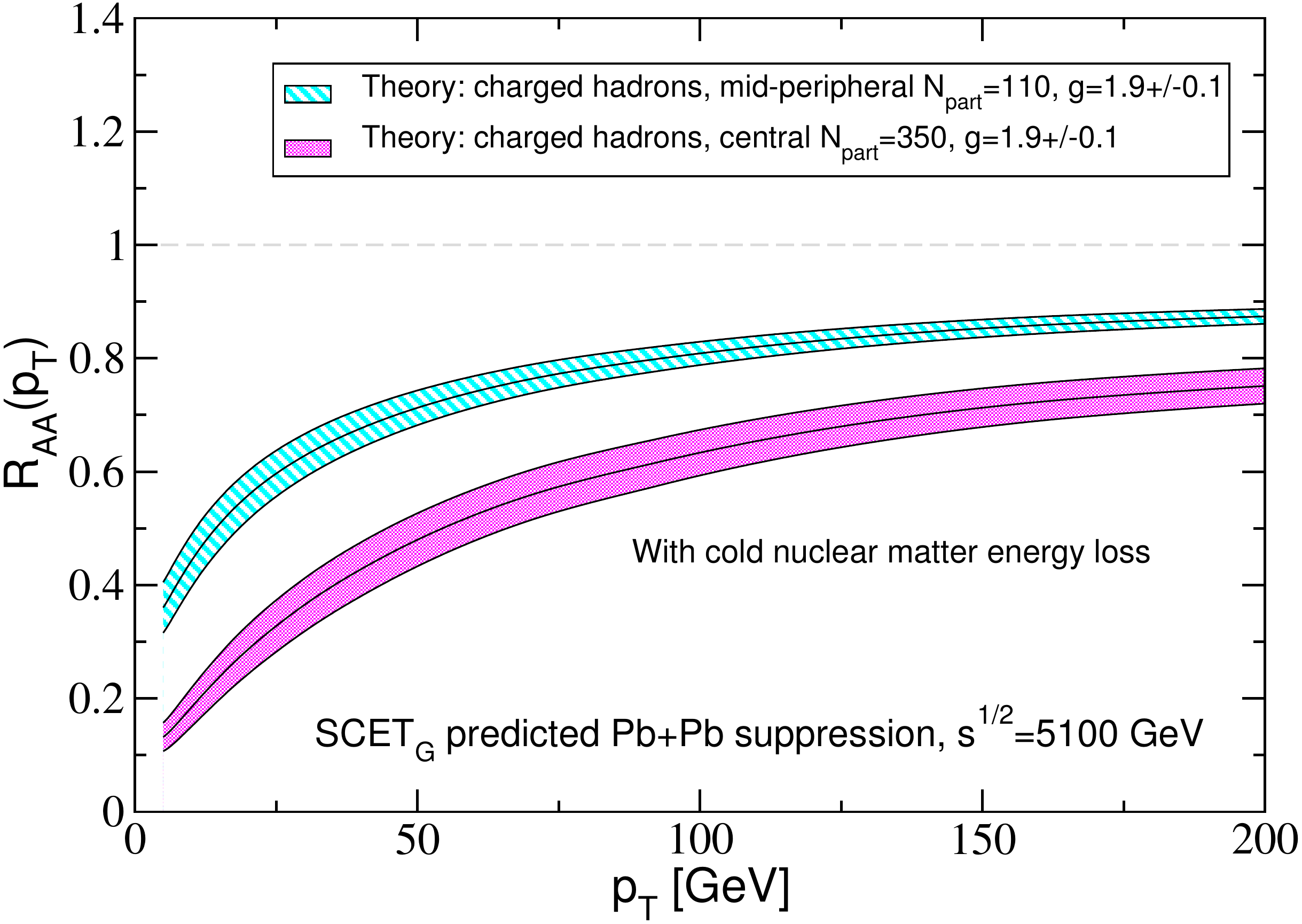} \ \ 
\includegraphics[width=7.4cm]{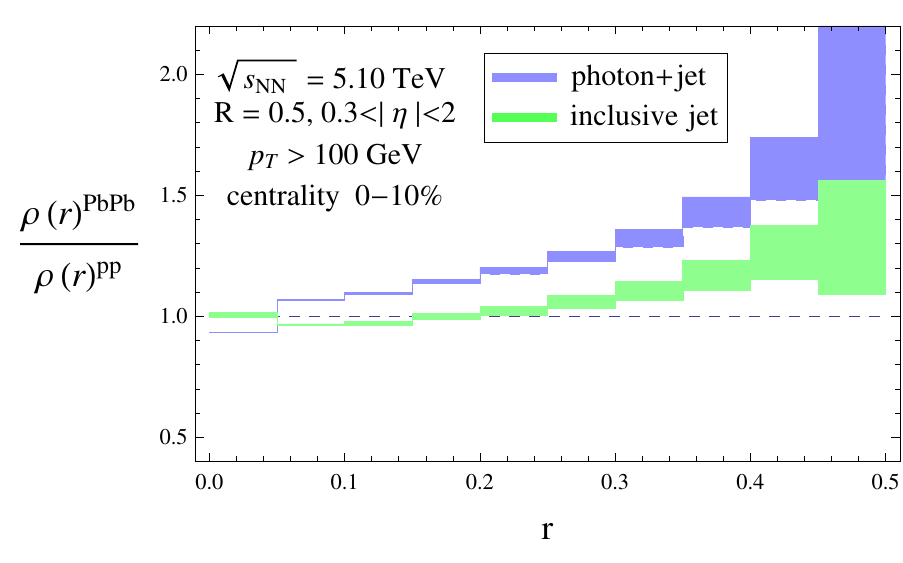}
\vspace*{-0mm}
\caption{Left panel: theoretical predictions for charged hadron $R_{AA}$ in central and mid-peripheral Pb+Pb  collisions at $\sqrt{s}\approx 5.1$~ATeV.  Right panel: related modification of the differential jet shapes for inclusive jets (green band) and photon-tagged jets (blue band), with $R=0.5$ in central Pb+Pb collisions. 
}\label{fig:pred}
\vspace*{-3mm}
\end{center}
\end{figure}

In summary, we presented results for the suppression of inclusive hadron production in Au+Au reactions at RHIC and Pb+Pb reactions at the LHC  based upon QCD factorization and DGLAP evolution with $\rm SCET_G$-based medium-induced splitting kernels. This method allowed us to unify the treatment of vacuum and medium-induced parton showers. In the soft gluon bremsstrahlung limit, we demonstrated the connection between this new approach and the traditional energy loss-based jet quenching phenomenology.  With an emphasis on a consistent theoretical descriptions of hadron and jet observables in heavy ion collisions,  we further  related the in-medium modification of parton showers to the attenuation of reconstructed jet cross sections and the modification of the jet shapes in heavy ion collisions. While good description of the majority of current experimental measurements is achieved, this theoretical framework can be further tested by the upcoming LHC Run II measurements. We present  predictions for charged hadron and neutral pion $R_{AA}$,  the jet shape modification and the cross section suppression for inclusive jets and photon-tagged jets see Fig.~\ref{fig:pred}. We fond that the cross section suppression at high $p_T$ can provide information about the cold nuclear matter effects.  Since photon-tagged jets are predominately quark-initiated, the cross section is expected to be less suppressed compared to inclusive jets. On the other hand, the broadening of the photon-tagged jet is more apparent.






\end{document}